\documentclass[aps,prl,floatfix,superscriptaddress,twocolumn,showpacs]{revtex4}

\usepackage{epsfig}
\usepackage{graphicx,psfrag}
\usepackage{dcolumn}
\usepackage{bm}

\newcommand{\beq}{\begin{equation}}
\newcommand{\eeq}{\end{equation}}
\newcommand{\bea}{\begin{eqnarray}}
\newcommand{\eea}{\end{eqnarray}}
\newcommand{\hf} {\frac{1}{2}}

\newcommand{\nn}{\nonumber\\}

\newcommand\eqn[1]     {Eq.\,(\ref{#1})}
\newcommand\eqns[2]    {Eqs.\,(\ref{#1}) and~(\ref{#2})}

\newcommand\fig[1]     {Fig.\,{\ref{#1}}}

\def\tu{{\tilde u}}

\def\ord#1{{\cal O}(#1)}

\def\mr#1{{\mathrm{#1}}}

\begin{document}

\title{Functional renormalization group approach to the sine-Gordon model}

\author{S. Nagy}
\affiliation{Department of Theoretical Physics, University of Debrecen,
Debrecen, Hungary}

\author{I. N\'andori}
\affiliation{Institute of Nuclear Research, P.O.Box 51, 
H-4001 Debrecen, Hungary} 

\author{J. Polonyi}
\affiliation{Strasbourg University, CNRS-IPHC, BP28 67037 Strasbourg Cedex 2, France}

\author{K. Sailer}
\affiliation{Department of Theoretical Physics, University of Debrecen,
Debrecen, Hungary}

\begin{abstract} 
The renormalization group flow is presented for the two-dimensional
sine--Gordon model within the framework of the functional renormalization
group method by including the wave-function renormalization constant.
The Kosterlitz--Thouless--Berezinski type phase structure is recovered
as the interpolating scaling law between two competing IR attractive
area of the global renormalization group flow.
\end{abstract}

\maketitle

{\it I.~Introduction.---}
The two-dimensional (2D) sine-Gordon (SG) model, defined by the bare action
\beq\label{sgact}
S =\int_x \left[\hf (\partial_{\mu}\varphi)^2 + u \cos(\beta\varphi) \right]
\eeq
in Euclidean spacetime, has already received a considerable amount of
attention \cite{Coleman,sine-G,Amit,Kehrein,Gersdorff,Nandori_SG,Nagy_SG}
since it is the simplest non-trivial quantum field theory with compact
variables. This feature is common with non-Abelian gauge theories and
is supposed to be the key to their confinement mechanism. In two dimensions
this is the driving force to form a non-trivial phase structure.
The SG model is known to belong to the universality class of the 2D
Coulomb gas and the 2D--XY spin model which have received important
applications in condensed matter systems, e.g. describing the
Kosterlitz--Thouless--Berezinski (KTB) \cite{KTB} phase transition of
vortices in a thin superfluid film. There is a continuous interest in the
literature in constructing SG type models to understand better the vortex
dynamics of condensed matter systems \cite{Benfatto}.

A more detailed relation between the SG model and the XY model in the
Villain-approximation is obtained by using lattice regularization \cite{Huang}.
The kinetic energy is periodic with the same period length as the potential energy.
Therefore the model supports vortices and has a third adjustable parameter, the 
vortex fugacity $z$. For $z\to0$ the vortices are suppressed and the SG model of
\eqn{sgact} is recovered in the continuum. The duality transformation,
$(\beta,u,z)\to(2\pi/\beta,2z,u/2)$ maps the continuum SG model ($z=0$) into the 
XY model without external field ($u=0$). The Coleman point \cite{Coleman}, 
separating the renormalizable, asymptotically free phase ($\beta^2<8\pi$)
and the non-renormalizable phase ($\beta^2>8\pi$) of the SG model is
mapped into the KTB point of the XY model.

The perturbative RG results beyond the local potential approximation (LPA)
\cite{Amit} can account for the KTB phase transition and provide $\beta^2\to 0$
for $\beta^2<8\pi$ in the infrared (IR) limit. Recently, by using the flow 
equation approach \cite{Kehrein} a different IR limit is obtained for
the frequency, i.e. $\beta^2\to 4\pi$. However, the latter method
is not able to recover the leading order  perturbative UV results for
$\beta^2<4\pi$, due to the wrong sign of the evolution equation derived for 
the frequency. Functional RG approaches have also been used to map the phase
structure of the SG model but their description is not complete since on the 
one hand the  LPA is used \cite{Nandori_SG,Nagy_SG} and on the other hand, 
the SG model is mapped onto other models belonging to the same universality class 
\cite{Gersdorff}. Therefore, the analysis of the SG model is still incomplete.

Our aim with this work is to determine the complete phase structure of the original
SG model by extending the functional RG analysis beyond the LPA, by including the
field-independent wave-function renormalization, as well. We use the functional
RG method for the effective average action \cite{Wetterich,EffRG,CS}
which enables us to treat the wave-function renormalization. The evolution
arises as the result of the gradual turning on of the field fluctuations
according to their increasing amplitude by decreasing the control
parameter $k$ from the initial value $\Lambda\ll k_0$ (with $k_0$
the UV cut-off which goes to infinity) to zero.

The phase structure is found to be the global result of a competition
between an IR fixed line and an IR fixed point. The traditional KTB
scaling law is actually an interpolation between these two effects.

{\it II.~The sine-Gordon model.---}
The functional renormalization group equation for the effective action of an 
Euclidean field theory  is \cite{Wetterich}
\beq\label{feveq}
k\partial_k\Gamma_k=\hf\mr{Tr}\frac{k\partial_k R_k}{R_k+\Gamma''_k}
\eeq
where the notation  $^\prime=\partial/\partial\varphi$ is used and the trace Tr
stands for the integration over all momenta. We use a power-law type regulator function
\beq
R_k = p^2\left(\frac{k^2}{p^2}\right)^b
\eeq
with the parameter  $b\ge 1$. \eqn{feveq} has been solved over the functional
subspace defined by the ansatz
\beq\label{eaans}
\Gamma_k = \int_x\left[\frac{z}2 (\partial_\mu\varphi_x)^2+V_k(\varphi_x)\right],
\eeq
with the local potential $V_k(\varphi)=\sum_{n=1}^\infty u_n(k)\cos(n\varphi)$
and the field-independent wave-function renormalization $z(k)$. \eqn{feveq} leads
to the evolution equations \cite{Nagy_MSG}
\bea\label{ea_v}
\partial_k V_k &=& \hf\int_p{\cal D}_kk\partial_k R_k,\\
\label{ea_wf}
k\partial_kz &=&
{\cal P}_0 V'''^2_k\int_p{\cal D}_k^2k\partial_k R_k\left(
\frac{\partial^2{\cal D}_k}{\partial p^2\partial p^2}p^2
+\frac{\partial{\cal D}_k}{\partial p^2}
\right)
\eea
with ${\cal D}_k=1/(zp^2+R_k+V''_k)$ and ${\cal P}_0=(2\pi)^{-1}\int_0^{2\pi} d\varphi$ 
being the projection onto the field-independent subspace.

{\it III.~Linearized scaling at the Coleman point.---}
We assume $\Lambda^2{\gg} k^2{\gg}|V_k''|$, keep the leading order terms in 
$V_k''$ in the Taylor-expansion of the r.h.s of \eqns{ea_v}{ea_wf} and
retain a single Fourier mode in the potential $V_k$ for simplicity.
The two-dimensional momentum integrals can easily be performed, giving
\bea
\label{u1lin}
(2+k\partial_k)\tu_1 &=& \frac1{4\pi z}\tu_1, \\
\label{zlin}
k\partial_kz &=& -\frac{\tu_1^2}{z^{2-2/b}}c_b,
\eea
where the dimensionless couplings $\tu_n = k^{-2} u_n$ are introduced, and
\beq
c_b = \frac{b}{48\pi}\Gamma\left(3-\frac2{b}\right)
\Gamma\left(1+\frac1{b}\right).
\eeq
The UV evolution \eqns{u1lin}{zlin} clearly show that the critical
value $z^*=1/8\pi$ at the Coleman point is independent of the blocking parameter $b$.
Furthermore the sharp cutoff limit $b\to \infty$ gives infinite value in the r.h.s.
of \eqn{zlin} signalling the impossibility of introducing the wave-function 
renormalization in that case. The RG trajectories obtained by integrating \eqns{u1lin}{zlin},
\bea\label{uzlin}
\tu_1^2(z) = \frac2{(8\pi)^{1-2/b} c_b}(z-z^*)^2+\tu_1^{*2},
\eea
indicate turning points in the vicinity of the fixed point, at 
$(\tu_1^\star=\tu_1(z^*), z^*)$. Such a flow exhibits the well-known 
features of the KTB type phase transition. Actually we see the dual of that 
transition as explained in the Introduction.

Thus \eqns{u1lin}{zlin} provide similar evolution around the KTB fixed point
as the one  already  obtained by a perturbative RG analysis \cite{Amit} and
the flow equation approach \cite{Kehrein} for the SG model, and also by
the real-space RG for the two-dimensional Coulomb gas \cite{Kosterlitz}. 
The KTB phase transition is characterized by the exponential dependence
of the correlation length on the inverse of the square-root of the reduced
temperature $t \propto \tu_1^{*2}$. The correlation length $\xi$ can be read
off from the scale $k^*\sim 1/\xi$  where the RG trajectories show up their
turning points. Inserting back the solution (\ref{uzlin}) into \eqn{zlin}
one obtains
\beq\label{crit}
\xi\sim e^{\sqrt{\pi}/(\tu_1^*8\sqrt{c_b})
+\tu_1^*(b{-}1)(2b{-}1)\sqrt{c_b}2^{1-6/b}\pi^{5/2-2/b}b^{-2}
+\ord{\tu_1^{*2}}
}
\eeq
which is the typical scaling law for KTB type phase transitions, modified
by analytic corrections vanishing for $\tu_1^*\to 0$. It is worthwhile
mentioning that only the quantitative details depend on the choice of the
parameter $b$ in the formula (\ref{crit}). Using \eqns{u1lin}{zlin} the
critical exponent $\eta$ can also be calculated via the vortex-vortex
correlation function \cite{Kosterlitz} and it is proved to take the value
$\eta=1/4$ independently of the parameter $b$.

{\it IV.~Coleman point, revisited as the dual KTB point.---} Let us
now take into account the higher-order terms of the Taylor expansion in $V_k''$,
as well as the higher harmonics of the local potential \cite{Nagy_SG}.
We choose $b=1$, corresponding to the Callan-Symanzik RG scheme \cite{CS}
which is free of UV divergences for $d=2$ and ultralocal. The evolution equations
assume a simpler form rendering easier the handling of the higher Fourier modes. 

The Fourier transform of \eqns{ea_v}{ea_wf} produces a set of coupled equations
for $\tu_n$ \cite{Nandori_SG} and $z$. We refer to the solution of these
equations with 10 Fourier modes as the {\it full} solution. According to our
experience the retaining of more Fourier harmonics modifies the flow in a
negligible manner. By restricting the solution to a single Fourier mode one
obtains the evolution equations
\bea
\label{u1full}
(2+k\partial_k)\tu_1 &=& \frac1{2\pi\tu_1 z} \left[1-\sqrt{1-\tu^2_1}\right],\\
\label{zfull}
k\partial_k z &=& -\frac1{24\pi}\frac{\tu_1^2}{(1-\tu_1^2)^{\frac32}},
\eea
whose solution will be referred as {\it exact in $\tu_1$}. The first two
terms in the Taylor expansion of \eqn{u1full} can be identified with the
approximation used in \cite{Amit} by a proper transformation of the parameters
in Callan-Symanzik RG scheme. The higher-order terms make negligible effect
on the evolution even in the neighbourhood of the turning point $(\tu_1^*, z^*)$
of the RG trajectory. We refer to the solution of \eqns{u1lin}{zlin} as the
{\it linearized} solution.
\begin{figure}[ht] 
\begin{center} 
\epsfig{file=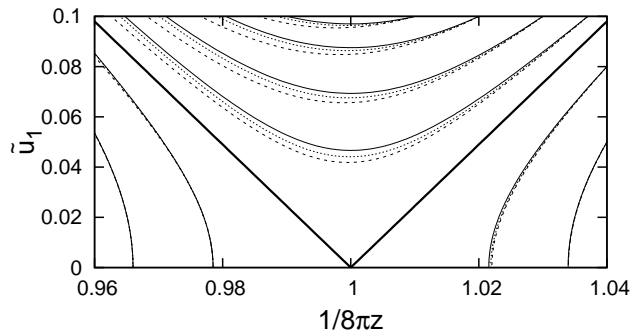,width=4.3cm,angle=-90}
\caption{\label{fig:phase}
Phase diagram of the SG model. The solid/ dashed/dotted lines show the RG trajectories for 
the linearized/exact in $\tu_1$/full solutions, respectively. The wide solid line depicts the
separatrix.
} 
\end{center}
\end{figure}
The RG trajectories are plotted in \fig{fig:phase}, they move to the left
as $k$ is decreased. This picture is reminiscent of the usual KTB phase structure.
What we see here is actually the vicinity of the dual of the KTB point of
the XY model \cite{Huang}. One can see that the higher harmonics modify the RG
trajectories rather slightly as compared to both the linearized and the exact in
$\tu_1$ solutions. Although the Coleman point lies at a crossover
scale between the UV and IR scaling regions for the trajectories
above the separatrix, the UV scaling remains valid in its vicinity. 

Note the smallness of the $z$-interval covered. The $z$-dependence of \eqn{ea_v}
is weak under the two solid lines of the separatrix, boardering three
different regions, where the regions from the left to the right correspond to
the renormalizable, the non-renormalizable and the asymptotically free regimes of
the SG model, respectively.

{\it V.~Non-perturbative scaling at the crossover.---}
Let us now turn our attention to the new phase which is opened up by the
evolution of the wave function renormalization constant in the middle
of the figure, above the separatrix. The $z$-dependence is crucial here,
it prevents the system to come to standstill where the renormalized
trajectory is stationary in $\tu_1$, just above the Coleman point.
It has been established that all Fourier modes are irrelevant (decrease with $k$)
before the crossover and they turn to relevant (starts to increase as
$k$ is further decreased) at $k^*_n$, showing very weak $n$-dependence at
the location of the turning point, $k^*_n\approx k^*$. We find
$z(k^*)=z^*=1/(8\pi)$, in a manner similar to the case $z=1$ \cite{Nagy_SG}.

The vertical line $z=z^*$ appears to be a single IR stable fixed 
point as far as the evolution of the potential is considered only.
In fact, the values of the coupling constants, $\tu_n^*=\tu_n(k^*)$,
determined by the Fourier transform of the evolution equation \eqn{ea_v} satisfy
at this line the condition that the ratios
\beq
c_n=\frac{\tu_n^*}{\tu_1^{*2n}},
\eeq
are universal constants, $c_2=1/12$, $c_3=1/96$, $c_4=13/8640$, $c_5=97/414720$,
etc. \cite{Nandori_SG}. We recover renormalizability and asymptotical freedom
in this phase because the dynamics is characterized by a single coupling strength,
$\tu_1>0$ at and below the crossover scale. This is a non-perturbative phenomenon
because the crossover "fixed point" is not Gaussian.

We see furthermore the subtle meaning of the "KTB fixed point". As soon as one
goes beyond the LPA the Coleman point ceases to be a fixed point and is
separating different phases only under the separatrix of \fig{fig:phase},
where the beta functions have a common analytic structure \cite{Nagy_SG} and something
irregularity shows up in the deep IR region of the symmetry broken phase only.

\begin{figure}[ht]
\begin{center}
\epsfig{file=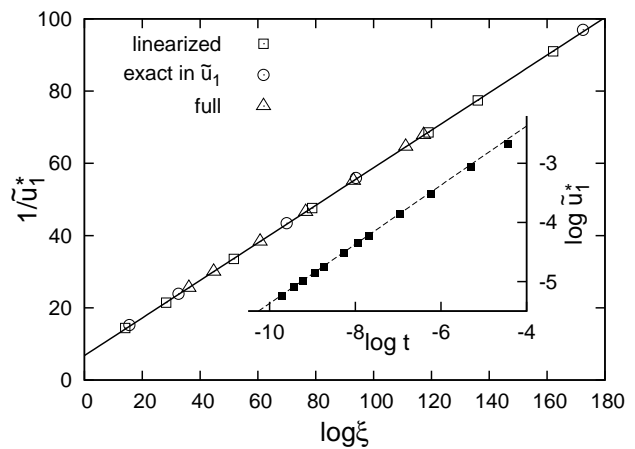,width=5.8 cm,angle=-90}
\caption{\label{fig:corr}
The inverse of the turning value $\tu_1^*$ of the fundamental amplitude is plotted
versus the  correlation length $\xi\sim1/k^*$ for
the linearized, exact in $\tu_1$ and full solutions, denoted by different
point types. The slope of the solid line is 
$\sqrt{8/3\pi^2}\approx 0.52$ according to \eqn{crit}. In the inset
the turning value $\tu_1^*$ is plotted against
the reduced temperature $t$. The slope of the
fitted dashed line is $0.5$.
}
\end{center}
\end{figure}

A more detailed and explicit similarity with the KTB scaling of the XY model is
found by introducing the correlation length $\xi$ by identifying it with the
inverse cutoff at the crossover. The numerical results are shown in \fig{fig:corr}.
The various approximations, i.e. the full solution, the solution exact in $\tu_1$,
and the linearized one give the same critical behaviour as \eqn{crit}, showing that
neither the inclusion of higher-order terms in \eqns{u1lin}{zlin} nor that of
the higher harmonics affect the type of the phase transition, their effects
are negligible. The reduced temperature is given formally by the wave-function
renormalization constant at the UV cutoff as $t=[z(\Lambda)-z_s(\Lambda)]/z_s(\Lambda)$,
where $(1/8\pi z_s(\Lambda),\tu_1^*(\Lambda))$ is a point of the separatrix.
The turning point $\tu_1^*$ is shown in the inset of \fig{fig:corr} as the function
of the reduced temperature $t$ for the linearized solution, giving 
\beq\label{uvst}
\tilde u_1^{*2}= qt+\ord{t^2}
\eeq
as in the XY model \cite{KTB}, and in the Coulomb gas \cite{Kosterlitz}. The same
relation is recovered for the exact solution in $\tu_1$ and for the full solution,
as well.
The critical scaling relations (\ref{crit}) and (\ref{uvst}) signal directly that there is a
KTB type phase structure in the SG model.

{\it VI.~The IR scaling regime.---}
The trajectories end in a line of Gaussian IR fixed point in the non-renormalizable phase.
All coupling strengths of the potential are irrelevant and this implies that the evolution
of the wave-function renormalization $z$ is extremely weak. The LPA can be used and the
well-known IR scaling is recovered, including the unusual feature of the non-availability
of the concept of relevant or irrelevant operators \cite{Nagy_SG}. There is a line of
Gaussian fixed points in the asymptotically free phase, too, but these fixed points are
UV and their scaling laws are linearizable. The IR scaling is difficult to establish
numerically because of the instability of the Fourier expansion
\cite{Nandori_scheme} in any RG scheme, used so far. Nevertheless it is 
unambiguous from numerics that $z$ tends to be big as the scale $k$ is decreased, 
while $\tu_1$ remains finite. 
\begin{figure}[ht]
\begin{center}
\epsfig{file=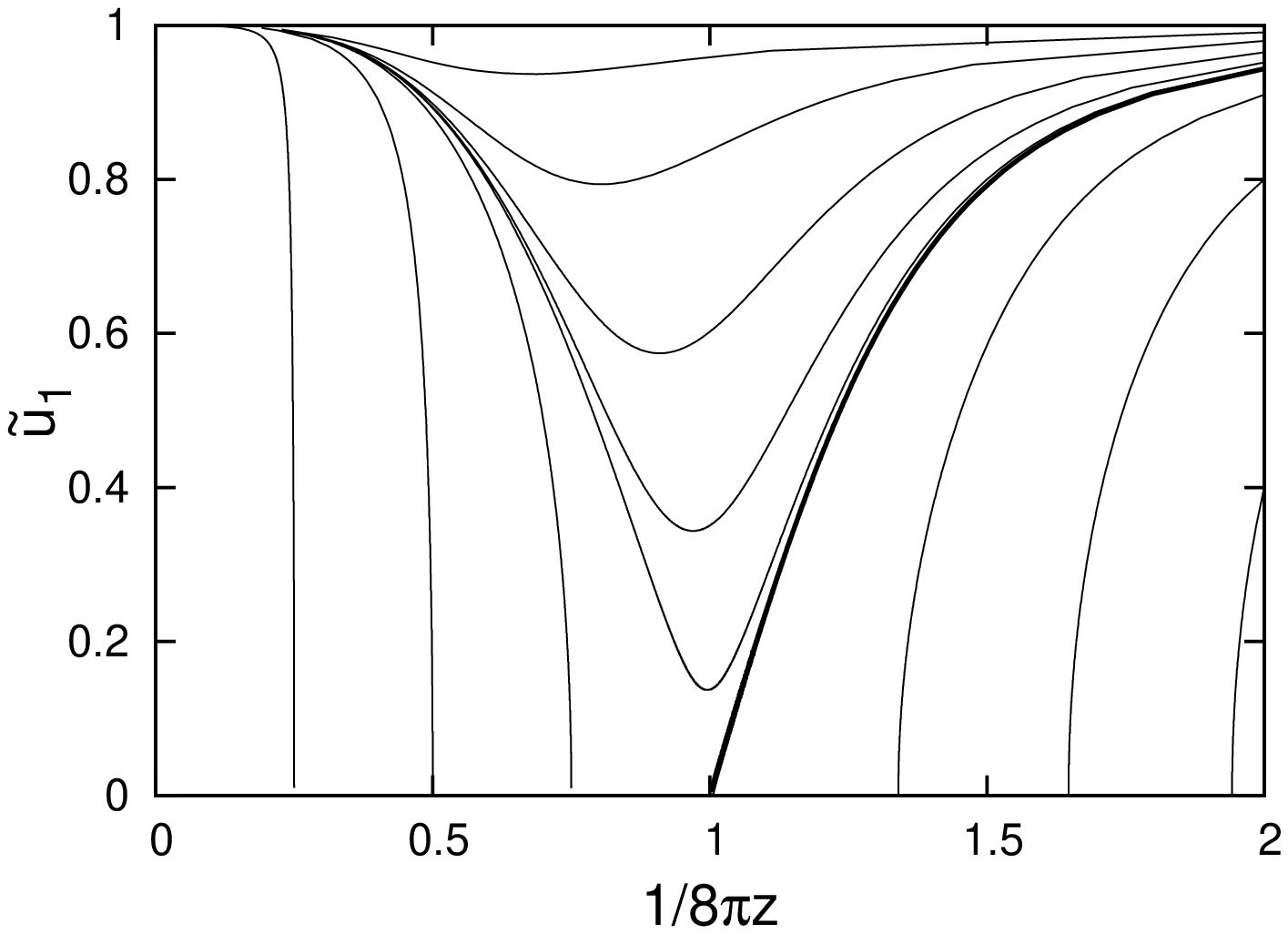,width=5.8 cm,angle=-90}
\caption{\label{fig:exph}
The phase structure of the SG model given by the numerical solution of \eqns{u1full}{zfull}.
}
\end{center}
\end{figure}
One can get a clear picture of the phase structure of the IR scaling regime by
omitting the effect of the higher harmonics, a frequently used approximation
\cite{Amit,Kehrein}. After introducing $\omega = \sqrt{1-\tu^2}$, $\chi=1/z\omega$
and $\partial_t = \omega^2 k\partial_k$ we arrive at the evolution equations
\bea
\partial_t\omega &=& 2\omega(1-\omega^2)-\frac{\omega^2\chi}{2\pi}(1-\omega),\nn
\partial_t\chi &=& \chi^2\frac{1-\omega^2}{24 \pi}-2\chi(1-\omega^2)+\frac{\omega\chi^2}{2\pi}(1-\omega),
\eea
possessing two lines of Gaussian fixed points separated by the well-known  
Coleman (alias KTB) fixed point, $(1/8\pi z,\tu_1)=(1,0)$, and an additional (IR)
fixed point $(1/8\pi z,\tu_1)=(0,1)$ (see  \fig{fig:exph} for the RG trajectories).

Such a modification of the scaling laws which is believed to preserve the qualitative
features of the RG flow makes the IR fixed points explicit in the complete phase diagram.
It also demonstrates that the hyperbolic nature of the flow in the vicinity
of the KTB-Coleman point stems from global effects, the competition between two
regions of the phase diagram. This is the attraction of the line of Gaussian IR fixed
points of the symmetrical phase, dominated by the kinetic energy on the one
hand and of the non-Gaussian, IR fixed point which is dominated by the
potential energy on the other.

The feature lost in this approximation is that the effective potential, built up
in the IR region makes the evolution equation singular, which is typical for
phases with spontaneously broken symmetry \cite{Tetradis}, automatically
guaranteeing the superuniversality for the potentials. In fact, the precise
treatment with no expansion should give the superuniversal potential
$\tilde V_{k\to 0} = -\hf\phi^2$ due to the Maxwell cut
\cite{Alexandre,Wetterich_kink,Nagy_SG} and $1/z(k\to 0) = 0$.


\begin{thebibliography}{99}
\bibitem{Coleman}
S. R. Coleman Phys. Rev. D {\bf 11}, 2088 (1975).
\bibitem{sine-G}
S. Mandelstam, Phys. Rev. D {\bf 11}, 3026 (1975);
J. V. Jos\'e, L. P. Kadanoff, S. Kirkpatrick, and D. R. Nelson Phys. Rev. B {\bf 16}, 1217 (1977);
B. Schroer, T. Truong, Phys. Rev. D {\bf 15}, 1684 (1977);
S. Samuel, Phys. Rev. D {\bf 18}, 1916 (1978);
AI. B. Zamolodchikov, Int. J. Mod. Phys. A {\bf 10}, 1125 (1995);
J. Zinn-Justin, {\it Quantum Field Theory and Critical Phenomena},
(Clarendon Press, Oxford, 1996).
\bibitem{Amit}
D. J. Amit, Y. Y. Goldschmidt and G. Grinstein, J. Phys {\bf A}13, 585 (1980);
P. B. Wiegmann, J. Phys C{\bf 11}, 1583 (1978);
J. Balog, A. Hegedus, J.Phys. A{\bf 33}, 6543 (2000).
\bibitem{Kehrein}
S. Kehrein, Phys. Rev. Lett. {\bf 83}, 4914 (1999).
\bibitem{Gersdorff}
G. v. Gersdorff, C. Wetterich, Phys. Rev. B {\bf 64}, 054513 (2001).
\bibitem{Nandori_SG}
I. N\'andori, J. Polonyi, K. Sailer Phys. Rev. D{\bf 63}, 045022 (2001).
\bibitem{Nagy_SG}
S. Nagy, I. N\'andori, J. Polonyi, K. Sailer, Phys. Lett. B{\bf 647}, 152 (2007).
\bibitem{KTB}
V. L. Berezinskii, Zh. Eksp. Teor. Fiz. {\bf 61}, 1144 (1971) [Sov. Phys.-JETP {\bf 34}, 610 (1972);
J. M. Kosterlitz, D. J. Thouless, J. Phys. C{\bf 6}, 1181 (1973).
\bibitem{Benfatto}
L. Benfatto, C. Castellani, T. Giamarchi, Phys. Rev. Lett. {\bf 98}, 117008 (2007);
Phys. Rev. Lett. {\bf 99}, 207002 (2007);
I. Nandori, S. Nagy, K. Sailer, U. D. Jentschura, Nucl. Phys B{\bf 725}, 467 (2005);
I. Nandori, K. Vad, S. Meszaros, U. D. Jentschura, S. Nagy, K. Sailer,
J. Phys.: Condens. Matter 19 (2007) 496211; ibid. 19 (2007) 236226.
\bibitem{Huang}
K. Huang, J. Polonyi, Int. J. of Mod. Phys. {\bf 6}, 409 (1991). 
\bibitem{Wetterich}
C. Wetterich, Phys. Lett. B{\bf 301}, 90 (1993).
\bibitem{EffRG}
J. Polchinski, Nucl. Phys B {\bf 231}, 269 (1984);
T. R. Morris, Int. J. Mod. Phys. A {\bf 9}, 2411 (1994);
D. F. Litim, Phys. Lett. B {\bf 486}, 92 (2000);
J. Polonyi, Central Eur. J. Phys.{\bf 1}, 1 (2004);
J. Comellas, Nucl. Phys. B{\bf 509}, 662 (1998);
M. E. Fisher, Rev. Mod. Phys. 70, 653 (1998);
C. Bagnuls, C Bervillier, Phys. Rept. {\bf 348}, 91, (2001);
J. Berges, N. Tetradis, C. Wetterich, Phys. Rept. {\bf 363}, 223 (2002).
\bibitem{CS}
J. Alexandre, J. Polonyi, Annals Phys. {\bf 288}, 37 (2001);
J. Alexandre, J. Polonyi, K. Sailer, Phys. Lett. B {\bf 531}, 316 (2002).
\bibitem{Nagy_MSG}
S. Nagy, I. N\'andori, J. Polonyi, K. Sailer, Phys. Rev. D{\bf 77}, 025026 (2008).
\bibitem{Kosterlitz}
J. M. Kosterlitz, J. Phys C{\bf 7}, 1046 (1974).
\bibitem{Nandori_scheme}
I. Nandori, S. Nagy, K. Sailer, A. Trombettoni, arXiv: 0903.5524.
\bibitem{Tetradis}
N. Tetradis, C. Wetterich, Nucl. Phys. B {\bf 383}, 197 (1992).
\bibitem{Alexandre}
J. Alexandre, V. Branchina, J. Polonyi, Phys. Lett. B {\bf 445}, 153 (1999).
\bibitem{Wetterich_kink}
C. Wetterich, Nucl. Phys. B {\bf 352}, 529 (1991).
\end{thebibliography}
\end{document}